\documentclass{svjour3}                     
\smartqed  
\usepackage{graphicx}
\begin{document}

\title{Experimental activities in few-body physics\footnote{Presented at the 21st European Conference on Few-Body Problems in Physics, Salamanca, Spain, 30 August - 3 September 2010.}
}


\author{J.G. Messchendorp}


\institute{J.G. Messchendorp \at
              Kernfysisch Versneller Instituut (KVI), University of Groningen, Zernikelaan 25, 9747 AA Groningen, The Netherlands \\
              Tel.: +31-503633558\\
              Fax:  +31-503634003\\
              \email{messchendorp@kvi.nl}           
}

\date{Received: date / Accepted: date}

\maketitle

\begin{abstract}
Understanding the few-nucleon system remains one of the challenges
in modern nuclear and hadron physics. Observables in few-nucleon scattering 
processes are sensitive probes to study the two and many-body interactions 
between nucleons in nuclei. In the past decades, several facilities provided 
a large data base to study in detail the three-nucleon interactions below the 
pion-production threshold by exploiting polarized proton and deuteron beams and
large-acceptance detectors. Only since recently, the four-nucleon scattering
process at intermediate energies has been explored. 
In addition, there is a focus to collect data in the hyperon-nucleon sector, 
thereby providing access to understand the more general baryon-baryon interaction.
In this contribution, some recent results in the few-nucleon sector are discussed 
together with some of the preliminary results from a pioneering and exclusive 
study of the four-nucleon scattering process. Furthermore, this paper
discusses the experimental activities in the hyperon sector, in particular,
the perspectives of the hyperon program of PANDA.
\end{abstract}

\section{Introduction}
\label{intro}

Understanding the exact nature of the nuclear force is one of the
long-standing questions in nuclear physics. In 1935, Yukawa
successfully described the pair-wise nucleon-nucleon (NN) interaction
as an exchange of a boson~\cite{Yukawa}. Current NN models are mainly
based on Yukawa's idea and provide an excellent description of the
high-quality data base of proton-proton and neutron-proton
scattering~\cite{stoks94} and of the properties of the
deuteron. However, for the simplest three-nucleon system, triton,
three-body calculations employing NN forces clearly underestimate the
experimental binding energies~\cite{wiringa95}, demonstrating
that NN forces are not sufficient to describe the three-nucleon system
accurately. Some of the discrepancies between experimental data and
calculations solely based on the NN interaction can be resolved by
introducing an additional three-nucleon force (3NF). Most of the
current models for the 3NF are based on a refined version of
Fujita-Miyazawa's 3NF model~{\cite{fuji}}, in which a 2$\pi$-exchange
mechanism is incorporated by an intermediate $\Delta$ excitation of
one of the nucleons~{\cite{deltuva03II,coon01}}. More
recently, NN and three-nucleon potentials have become available 
which are derived from the basic symmetry properties of 
the fundamental theory of Quantum Chromodynamics (QCD)~\cite{chiptnn1,chiptnn2}.
These so-called chiral-perturbation ($\chi$PT) driven models construct systematically
a potential from a low-energy expansion of the most general
Lagrangian with only the Goldstone bosons, e.g. pions, as exchange
particles. The validity of the $\chi$PT-driven models for the intermediate energies
remains, however, questionable and depends strongly on
the convergence of results at higher terms in the momentum expansion.

\section{Nucleon-deuteron scattering}

\begin{figure}[ht]
\centering 
\includegraphics[width=0.8\textwidth]{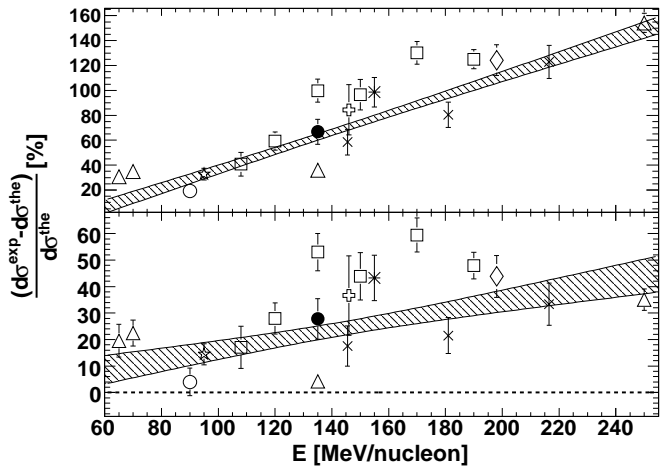}
\caption{The relative difference between the calculations by the Hannover-Lisbon 
theory group and the measured cross sections for the elastic $p+d$ reaction as a
function of beam energy for $\theta_{\rm c.m.}=140 ^\circ$. 
The top panel shows the differences with a calculation based
on the CD-Bonn potential and the Coulomb interaction, whereas for the 
bottom panel an additional $\Delta$ isobar has been taken into account.
Open squares are data from Ref.~{\cite{kars03}}, open triangles are data from
Refs.~\cite{sakai00,kimiko02,shimi,hatan}, open circle is from~{\cite{hamid}},
open star is from~{\cite{mermod}}, crosses are from~{\cite{Igo}}, star
is from~{\cite{kurodo}}, open cross is from~{\cite{postma}}, diamond is from
~{\cite{ald}} and the filled circle is from~{\cite{ram08}}.  The shaded
band represents the result of a line fit through
the data excluding the results obtained at KVI, RIKEN and RCNP. 
The width of the band corresponds to a 2$\sigma$ error of the fit.}
\label{syscheck} 
\end{figure}

In the last decade, high-precision data at intermediate
energies in elastic \it {Nd} \rm and \it {dN} \rm
scattering~{\cite{bieber,kars01,kars03,kars05,sakai00,kimiko02,kimiko05,postma,hamid,kurodo,mermod,Igo,ald,Hos,Ela07,shimi,hatan,IUCF}}
for a large energy range together with rigorous Faddeev
calculations~{\cite{gloeckle}} for the three-nucleon system have
proven to be a sensitive tool to study the 3NF. In particular, a large
sensitivity to 3NF effects exists in the minimum of the differential
cross section~{\cite{witala98,nemoto}}. 
The results of a systematic study of the energy dependence of 
all available cross sections in elastic proton-deuteron scattering 
with respect to state-of-the-art calculations by the Hannover-Lisbon 
theory group are depicted in Fig.~\ref{syscheck}. The top panel shows 
the relative difference between the model predictions 
excluding the $\Delta$-isobar contribution and data taken at a fixed 
center-of-mass angle of $\theta_{\rm c.m.}$=140$^\circ$. The data points 
were extracted from a polynomial fit through each angular distribution. 
The error bars correspond to a quadratic sum of the statistical and systematic
uncertainties of each measurement. Note that the discrepancies, reflecting
the 3NF effects, increase drastically with incident energy and reach 
values of more than 100\% at energies equal or larger than 200~MeV.
The bottom panel in Fig.~\ref{syscheck} shows a similar comparison
between data and model predictions including the $\Delta$-isobar as
mediator of the 3NF effects. Clearly, a large part of the discrepancies
is resolved. However, a smaller but significant deficiency remains which
increases with energy to values of about 30\% at an energy of 200~MeV.

Complementary to the elastic scattering experiments, three-nucleon
studies have been performed exploiting the nucleon-deuteron break-up
reaction. The phase space of the break-up channel is much richer than that of
the elastic scattering. The final state of the break-up reaction is
described by five kinematical variables, as compared to just one for the
elastic scattering case. Therefore, studies of the break-up reaction
offer a way of much more detailed investigations of the nuclear
forces, in particular of the role of 3NF effects. Predictions show that 
large 3NF effects can be expected at specific kinematical regions in the
break-up reaction. Results of the cross sections and tensor analyzing
powers have already been published for a deuteron-beam energy of 130~MeV 
on a liquid-hydrogen target~{\cite{kistryn06,ola06,ela07}}. These
experiments were the first ones of its type which demonstrated the 
feasibility of a high-precision measurement of the break-up observables and they
confirmed that sizable influences of 3NF and Coulomb effects are
visible in the break-up cross sections at this energy. In the last
years, more data at several beam energies and other observables have
been collected to provide an extensive data base at intermediate
energies.

\begin{figure}[ht]
\centering
\includegraphics[angle=0,width=0.7\textwidth]{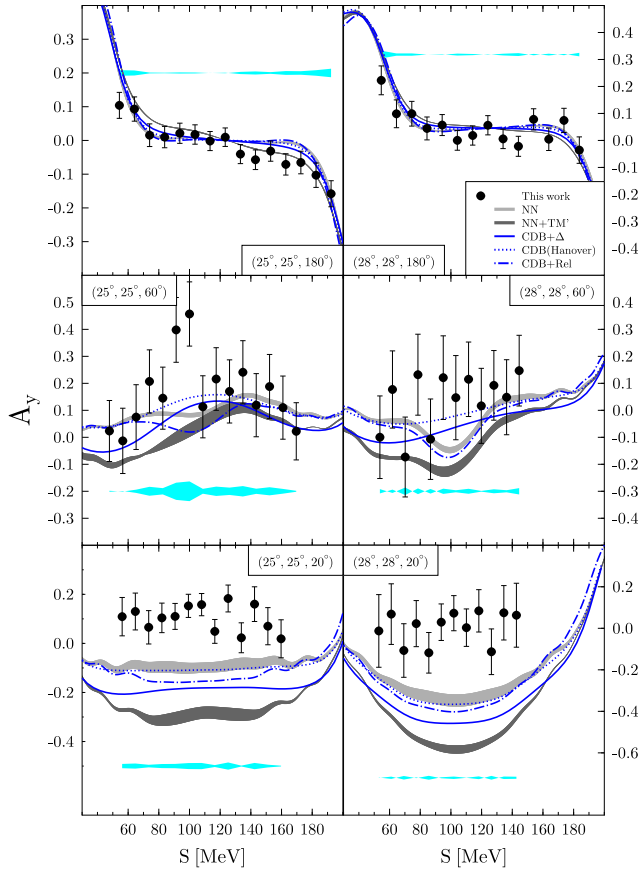}
\caption{A comparison between the results of the analyzing power
 measurements for a few selected break-up configurations with various
 theoretical predictions. The light gray bands are composed of various
 modern two-nucleon (NN) force calculations, namely CD-Bonn, NijmI,
 NijmII, and AV18. The dark gray bands correspond to results of the
 calculations with the same NN forces including the TM' (3N)
 potential.  The lines represent the predictions of calculations by
 the Hannover-Lisbon group based on the CD-Bonn potential (dotted) and
 CD-Bonn potential extended with a virtual $\Delta$ excitation (solid blue). 
 The blue dash-dotted lines are derived from calculations by
 the Bochum-Cracow collaboration based on the CD-Bonn potential
 including relativistic effects~\cite{skib06}. The errors are statistical and the
 cyan band in each panel represents the systematic uncertainties
 (2$\sigma$).}
\label{crossall}
\end{figure}

Recent and interesting results have been obtained at KVI using a 4$\pi$
detection system BINA, which provides a unique tool to study a large
part of the phase space of the break-up reaction. 
Figure~\ref{crossall} presents some results of the vector
analyzing powers in proton-deuteron break-up for an incident proton
beam of 190~MeV and for two symmetric kinematical configurations 
$(\theta_{1},\theta_{2})$=($25^{\circ},25^{\circ}$) and 
($28^{\circ},28^{\circ}$) for three different values of $\phi_{12}$. 
Here, the angles $\theta_1$ and $\theta_2$ refer to the polar angles
of the two final-state protons and $\phi_{12}$ to the relative azimuthal
angle between these protons. The parameter $S$ is directly related to the
energies of the two final-state protons and is a measure of their energy
correlation. The data are compared with
calculations based on different models for the interaction dynamics as
described in detail in the caption of the figure. For these configurations and
observable, the effects of relativity and the Coulomb force are
predicted to be small with respect to the effect of three-nucleon
forces. At $\phi_{12}$=$180^\circ$, the value of $A_y$ is predicted
to be completely determined by two-nucleon force effects with only a
very small effect of 3NFs, which is supported by the experimental
data. Note, however, that the effect of 3NFs increases with decreasing
of the relative azimuthal angle $\phi_{12}$, corresponding to a
decrease in the relative energy between the two final-state protons.
The observed discrepancies could point to a
deficiency in the spin-isospin structure of the description of the
many-nucleon forces in the present-day state-of-the-art calculations
as discussed in Ref.~\cite{mar09}.

\begin{figure}[ht]
\centering
\includegraphics[angle=0,width=0.7\textwidth]{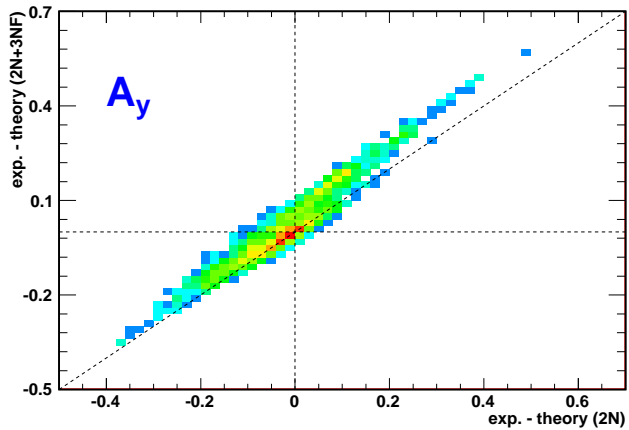}
\caption{A global analysis of the proton-deuteron break-up data on the vector analyzing
power, $A_y$, for two incident proton-beam energies of 135~MeV and 190~MeV. The figure
includes about 2500 data points for a large range in kinematics. The data are compared
with predictions by the Hannover-Lisbon theory group. For a more detailed description, see
text.}
\label{globalanalysis}
\end{figure}

A more global analysis is presently conducted in which the deficiencies with the
state-of-the-art calculations are being systematically studied for all the 
available break-up data. These results will soon become available within a 
review article. For this paper, I present one of the preliminary outcomes of this analysis.
In Fig.~\ref{globalanalysis} a summary plot is depicted for all the available
vector-analyzing data points in proton-deuteron break-up for two incident proton energies, 
namely 135~MeV and 190~MeV. The figure depicts the deviations
with respect to calculations from the Hannover-Lisbon 
theory group. The y-axis represents the deviation with a calculation based upon
the CD-Bonn potential and excluding the effect of the 3NF, whereas the x-axis
represents the deviation with the same calculation including 3NF effects. Here,
the $\Delta$ resonance within a coupled-channel framework mimics the 3NF effects.
The calculation takes into account the effect of the Coulomb force. The color intensity
on the z-axis represents the number of kinematical configurations, e.g. data points, 
that fall into the corresponding bin. Note that a large amount of data points are close
to the origin. For those cases, the calculations predict a negligible 3NF effect and 
the data agrees well with the model predictions. In the ideal case, one hopes that all 
the data would lie on the horizontal line, implying that the 3NF effects are correctly 
incorporated in the model. Data that fall on the diagonal line away from the origin would imply that 
the calculations predict only a small 3NF effect, whereas the data are incompatible
with this assumption. The horizontal line indicates the worse case scenario, e.g. 
the inclusion of a 3NF effect makes the discrepancy with the experimental data larger.
Strikingly, a large fraction of the break-up data fall within the diagonal and vertical
line, indicating that our present understanding of 3NF effects is not under control for
this channel and observable. 

\section{The next generation few-body experiments}


\begin{figure}[ht]
\centering
\includegraphics[width=0.7\columnwidth,angle=0]{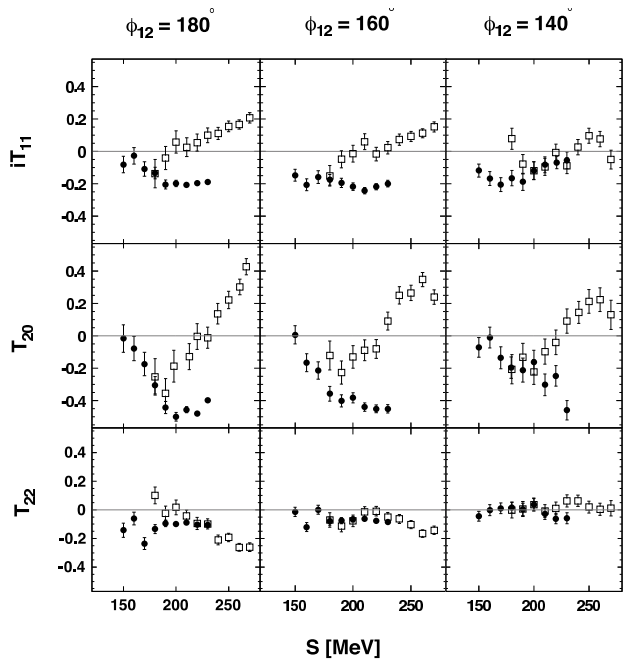}
\vspace*{-0.2cm}
\caption{The vector- and tensor-analyzing powers at $(\theta_{d},
  \theta_{p}) = (15^{\circ}, 15^{\circ})$ (open squares) and
  $(\theta_{d}, \theta_{p}) = (25^{\circ}, 25^{\circ})$ (filled
  circles) as a function of $S$ for different azimuthal opening
  angles. Only statistical uncertainties are indicated. The
  total systematic uncertainty is estimated to be \textbf{$\sim 7.5\%$}.}
\label{dd1}
 \end{figure}

The 3NF effects are in general small in the three-nucleon system. A
complementary approach is to look into systems for which the 3NF
effects are significantly enhanced in magnitude. For this, it was
proposed to study the four-nucleon system. The experimental data base
in the four-nucleon system is presently poor in comparison with the
three-nucleon system.  Most of the available data were taken at very
low energies, in particular below the three-body break-up threshold of
2.2~MeV. Also, theoretical developments are evolving rapidly at low
energies~\cite{RamazaniA_Ref23,RamazaniA_Ref24,RamazaniA_Ref25,RamazaniA_Ref26},
but lag behind at higher energies. The experimental data base at
intermediate energies is very
limited~{\cite{RamazaniA_Ref27,RamazaniA_Ref28,RamazaniA_Ref29}}.
This situation calls for extensive four-nucleon studies at
intermediate energies. 

Recently, comprehensive measurements of 
cross sections and spin observables in various $d+d$ scattering processes 
at 65~MeV/nucleon, namely the elastic and three-body break-up channels, 
were performed at KVI using the BINA detector. With the corresponding results, 
the four-nucleon scattering data base at intermediate energies is significantly enriched. 
Figure~\ref{dd1} depicts some of the preliminary results of the deuteron-deuteron
three-body break-up reaction, $d+d\rightarrow d+p+n$, which were obtained via the 
unambiguous detection of a proton in coincident with a deuteron in the final state. 
For the first time, a systematic and exclusive study of the three-body
break-up reaction in deuteron-deuteron scattering at intermediate energies
was shown to be feasible and provided precision results in the
four-nucleon sector as well.  


Although the experimental database in the two- and three-nucleon system is
very extensive, data on hyperon-nucleon and hyperon-hyperon interactions are
very scarce. Fundamentally, however, an experimental (in combination with a theoretical)
study of the interaction between strange-rich baryons is of highly interest for the 
few-body community, since it will allow to understand in more detail the
general interactions between baryons. Most of the data on hyperon-nucleon scattering
stems from bubble chamber studies from the 60's and 70's. Theoretically,
major progress has been made for these systems. For example, effective field
theoretical predictions are underway~\cite{pol06}. Clearly, the experimental
progress is presently lacking behind.

The most promising experimental studies will come from various hypernuclear 
experiments conducted worldwide. Various single $\Lambda$-hypernuclei have
been discovered already by using beams of kaons, pions, or virtual photons
to convert a nucleon to a hyperon inside the nucleus and by exploiting the
missing-mass technique. Combined with gamma detection, precision
spectroscopy studies can be performed to reveal the details of the hyper-nucleon
interaction.

A study of the hyperon-hyperon interaction becomes possible via spectroscopy
studies of $\Lambda\Lambda$ hypernuclei. The existence of these exotic hypernuclei
has been established decades ago, already in 1966~\cite{prow66}. So far, only a few 
$\Lambda\Lambda$ hypernuclei has been discovered, which leads to an enormous discovery 
potential for future experiments. The PANDA (antiProton ANnihilations at DArmstadt)
experiment at the future FAIR facility in Darmstadt, Germany, will provide an excellent
tool to study $\Lambda\Lambda$ hypernuclei with great precision.
The key is to make use of a beam of antiprotons which allows to produce 
$\Xi^-$ particles with a relatively large cross section. The $\Xi^-$ particles
are slowed down and captured in a second target nucleus where they can convert eventually
into two $\Lambda$ particles with the emission of a 28~MeV photon. The
$\Lambda\Lambda$ hypernuclei are sub-sequentially identified and studied via
gamma spectroscopy (using a half sphere of Germanium detectors) and via the weak decay products
of the hyperons. The feasibility of such 
experiment has been proven and reported in Ref.~\cite{pandapb}. 
The hypernuclear study will be an integral part of a much broader physics 
program by the PANDA collaboration. A detailed description of the complete physics
program of PANDA can be found in Ref.~\cite{pandapb}.

\section{Conclusions}

In the past decades, our understanding of the nuclear forces has
drastically improved. These developments can be attributed to the
enormous progress made in theory and in experiment. In particular,
in the three-nucleon sector, the theoretical descriptions are 
ab-initio, based on high quality potentials, and (partly) able
to include effects like Coulomb and relativity. Also, the
experimental techniques have significantly improved in the course
of time and have provided a huge data base with high-precision
data and covering a huge part of the phase space. The
four-nucleon data base at intermediate energies is growing
significantly, thereby providing potentially new insights 
and a testing ground for our present understanding of the 
many-body force effects.

In spite of the progress made in experimental and theoretical techniques to
study the many-nucleon system, there are still various open questions
which urgently need to be addressed. A large part of these questions
point to our present understanding of 3NF effects. This paper discusses
some results of few-nucleon scattering experiments taken at intermediate energies. 
Although, the overall comparison between data 
and theory improve significantly by taking into account 3NF effects, 
there are still various channels, phase spaces, and observables which 
show huge discrepancies. Therefore, the existing data base for few-nucleon
scattering observables provide an ideal basis to develop a better understanding of
three-nucleon force effects in few-nucleon interactions.

From an experimental point-of-view, the future ``few-body'' challenge would lie 
in providing accurate data in the hyperon-nucleon and hyperon-hyperon sector. 
With this, the hope is to enrich the study of the nucleon-nucleon force towards the 
more general baryon-baryon interaction. Future experiments, such as PANDA,
show good perspectives in this direction.

\begin{acknowledgements}
This work is part of the research program of the Stichting voor Fundamenteel Onderzoek 
der Materie (FOM) with financial support from the Nederlandse Organisatie 
voor Wetenschappelijk Onderzoek (NWO). The present work 
has been performed with financial support from the University of Groningen and the
GSI, Helmholtzzentrum f\"ur Schwerionenforschung GmbH, Darmstadt. 
\end{acknowledgements}


\end{document}